\documentstyle[11pt,paspconf]{article}

\begin{document}

\newcommand{\lx}{\mbox{$L_{\rm X}$}}
\newcommand{\lxlbol}{\mbox{$L_{\rm X}/L_{\rm bol}$}}
\newcommand{\ross}{\mbox{$N_{R}$}}
\newcommand{\vsini}{\mbox{$v \sin i$}}
\newcommand{\kms}{\mbox{km\,s$^{-1}$}}
\newcommand{\lii}{\mbox{Li\,{\sc i}}}
\newcommand{\ki}{\mbox{K\,{\sc i}}}

\title{Lithium depletion in open clusters}
\author{R. D. Jeffries}
\affil{Department of Physics, Keele University, Staffordshire
ST5 5BG, UK}

\begin{abstract}
The current status of observational studies of lithium depletion in
open clusters is reviewed, concentrating mainly on G and K type
stars. I {\em attempt} to answer the following questions: Can the
lithium depletion patterns seen in open clusters be explained in terms
of standard stellar evolution models? What is the observational
evidence for non-standard mixing processes and on what timescales do
they operate? Does metallicity play a significant role in lithium
depletion? Can lithium still be used as a means of dating young stars?
What future observations might yield better answers to these questions?
\end{abstract}
           

\section{Introduction}
Lithium is the only metal produced in significant quantities in the Big
Bang. In principle, measurements of Li in old Population II stars
yield the primordial Li abundance, which would (in conjunction with
$H_{0}$) strongly constrain the universal baryon density and Big Bang
nucleosynthesis models. Sadly, in addition to processes which 
create Li in the universe, there are mechanisms which lead to its
destruction in stellar interiors via $p,\alpha$ reactions at only
$(2-3)\times10^{6}$\,K. There is debate about whether the $A({\rm Li})$
($=12 + \log[N({\rm Li})/N({\rm H})]$) value of 2.1-2.2 measured in
Population II stars is almost undepleted from the primordial value, or
whether the primordial value is closer to the $A({\rm Li})$ of 3.3
measured in the youngest stars and solar system meteorites, and has been
significantly depleted in Population II stars (Bonifacio \& Molaro
1997, Deliyannis \& Ryan 1997).

The former interpretation requires processes that increase the Galactic
Li abundance by factors of 10 in $\simeq 5$\,Gyr, while the latter
requires us to rethink the way that material is mixed in stellar
interiors. {\em Standard models}, which incorporate only convective
mixing, predict little ($\simeq0.1$ dex) Li depletion in Population II
stars. However, many extensions to the standard model, incorporating
{\em non-standard} mixing such as microscopic diffusion, turbulence
induced by rotational instabilities, meridional circulation and
gravitational waves have been proposed and developed in some detail by
a number of groups (see Pinsonneault 1997 for a review).

Open clusters are excellent laboratories for investigating non-standard
stellar physics relating to Li depletion.  We can assume that we have
co-eval groups of stars with very similar compositions and by choosing
clusters with a range of ages and compositions we can hope to answer
the questions posed in the abstract. 
The observational database for Li
in open clusters has grown enormously in the last 10 years, thanks to
sensitive detectors and the accessibility and strength of the
\lii\,6708\AA\ resonance doublet upon which most abundance measurements
are based. Table~1 gives a summary of these observations, listing
clusters, ages and distances (from the Lyng\aa\ 1987 catalogue -- treat
with extreme caution!), the number and spectral-types of (main-sequence
[MS] or pre-main sequence [PMS]) stars surveyed, with references.

\begin{center}
\begin{table}
\scriptsize
\caption{Open clusters with late-type Li abundance measurements}
\begin{tabular}{lccrcl}
\hline
Cluster & Log Age & Distance & No. & Types & Refs. \\
        & (yr)& (pc)    & &  &       \\
\hline
IC 2602   & 7.00 & 155   & 25 &FGKM   & Randich et al. 1997, A\&A, 323,
                                        86; \\
IC 2602   &      &       & 26 &GKM    & Meola et al. these proceedings\\
NGC 2264  & 7.30 & 750   &  6 &FG     & King 1998, AJ, 116, 254;\\
          &      &       & 28 &GK     & Soderblom et al. 1999, AJ in press\\
IC 2391   & 7.56 & 140   & 10 &FGKM   & Stauffer et al. 1989, ApJ, 342, 285;\\
IC 2391   &      &       & 22 &GKM    & Meola et al. these proceedings\\
IC 4665   & 7.56 & 430   & 14 &GKM    & Mart\'{i}n \& Montes 1997, A\&A,
                                        318, 805\\
Blanco 1  & 7.70 & 190   & 39 &FGK    & Panagi \& O'Dell 1997, A\&AS,
                                        121, 213;\\
Blanco 1  &      &       & 17 &GK     & Jeffries \& James 1999, ApJ,
                                        511, 218\\
Alpha Per & 7.71 & 170   &  5 &F      & Boesgaard et al. 1988, ApJ,
                                        327, 389;\\
Alpha Per &      &       &  3 &M      & Garcia-Lopez et al. 1994, A\&A,
                                        282, 518;\\
Alpha Per &      &       & 29 &FGK    & Balachandran et al. 1996, ApJ,
                                        470, 1243;\\
Alpha Per &      &       & 18 &KM     & Randich et al. 1998, A\&A, 333, 591\\
NGC 2547  & 7.76 & 400   & 34 &KM     & Jeffries et al. these proceedings\\
Pleiades  & 7.89 & 125   & 17 &F      & Boesgaard et al. 1988, ApJ,
                                        327, 389;\\
Pleiades  &      &       & 95 &FGK    & Soderblom et al. 1993, AJ, 106, 1059;\\
Pleiades  &      &       & 13 &K      & Garcia-Lopez et al. 1994, A\&A,
                                        282, 518;\\ 
Pleiades  &      &       & 15 &KM     & Jones et al. 1996, ApJ, 112, 186;\\
Pleiades  &      &       &  8 &GK     & Russell 1996, ApJ, 463, 593\\
NGC 2516  & 8.03 & 440   & 24 &FGK    & Jeffries et al. 1998, MNRAS,
                                        300, 550\\
NGC 1039  & 8.29 & 440   & 34 &FGK    & Jones et al. 1997, AJ, 114, 352\\
NGC 6475  & 8.35 & 240   & 35 &FGK    & James \& Jeffries 1997, MNRAS, 292, 252\\
Coma Ber  & 8.60 &  86   & 16 &F      & Boesgaard 1987, ApJ, 321, 967; \\
Coma Ber  &      &       &  5 &FG     & Soderblom et al. 1990, AJ, 99, 595;\\
Coma Ber  &      &       & 15 &FGK    & Jeffries 1999, MNRAS, 304, 821\\
Coma Ber  &      &       & 11 &FGK    & Ford et al. 1999, A\&A submitted\\
NGC 6633  & 8.82 & 320   & 21 &FGK    & Jeffries 1997, MNRAS, 292, 177\\
Hyades    & 8.82 &  48   & 32 &F      & Boesgaard 1987, PASP, 99, 1067;\\
Hyades    &      &       & 14 &F      & Boesgaard \& Budge 1988, ApJ,
                                        332, 410;\\
Hyades    &      &       & 23 &FG     & Soderblom et al. 1990, AJ, 99, 595;\\
Hyades    &      &       & 68 &FGK    & Thorburn et al. 1993, ApJ, 415, 150\\
Hyades    &      &       & 12 &K      & Soderblom et al. 1995, AJ, 110,
                                        729\\
Hyades    &      &       &  7 &K      & Barrado \& Stauffer
                                        1996, A\&A, 310, 879\\ 
Praesepe  & 8.82 & 180   & 63 &FG     & Soderblom et al. 1993, AJ, 106, 1080\\ 
NGC 752   & 9.04 & 400   & 19 &FG     & Hobbs \& Pilachowski 1986, ApJ,
                                         309, L17;\\
NGC 752   &      &       &  6 &F      & Hobbs \& Pilachowski 1988,
                                        PASP, 100, 336\\
NGC 3680  & 9.26 & 800   & 16 &FG     & Pasquini et al. 1998, CSSS10, CD-947\\
M67       & 9.60 & 720   &  7 &FG     & Hobbs \& Pilachowski 1986, ApJ,
                                        311, L37;\\ 
M67       &      &       &  6 &F      & Spite et al. 1987, A\&A, 171, L1;\\ 
M67       &      &       & 14 &FG     & Pasquini et al. 1997, A\&A,
                                        325, 535;\\
M67       &      &       & 27 &FG     & Barrado et
                                        al. 1997, MSAI, 68, 939\\ 
M67       &      &       & 25 &FG     & Jones et al. 1999, AJ, 117, 330\\
NGC 188   & 9.70 & 1550  &  7 &F      & Hobbs \& Pilachowski 1988, ApJ,
                                        334, 734\\
\hline
\end{tabular}
\end{table}
\end{center}

\section{Models of Li Depletion}
Figure 2 in the review of Pinsonneault (1997) gives an
overview of the Li depletion predictions of standard stellar evolution
models, where convection (and some convective
overshoot) is the only mixing mechanism. Quantitatively, models
produced by various groups depend upon the details of the adopted convective
treatment and atmospheric opacities ({\em e.g.} D'Antona \& Mazzitelli 1994). 
Qualitatively, there is general
agreement that (a) In G and K stars Li depletion occurs mainly during
the PMS phase with hardly any depletion on the main sequence for stars
hotter than 5000\,K. This is caused by the growth of the radiative
zone, pushing the base of the convection zone (CZ) outward to temperatures
too cool to burn Li. 
(b) Li depletion is {\em strongly} dependent on metallicity.
A high metallicity leads to greater opacity, deeper CZs
with higher base temperatures and hence greater Li depletion. A 0.2 dex
change in mean metallicity should lead to an order of magnitude change
in the PMS Li depletion at 5000\,K.

Models incorporating non-standard mixing modes predict Li depletion
{\em in addition} to that provided by PMS convection. For instance,
Chaboyer, Demarque \& Pinsonneault (1995) show that in G and K stars
there is little extra depletion during the PMS phase and then depletion
continues during the MS phase driven by instabilities associated with
rotation and angular momentum loss. The faster rotators on the ZAMS are
predicted to have a higher depletion rate. There is also likely to be
some metallicity dependence as well, because the distance between the
CZ base and where Li can be burned will be important. Other mechanisms
such as microscopic diffusion are expected to be much less important in
these stars with relatively deep convective envelopes.

\section{The Pleiades and Hyades}

\begin{figure}
\vspace*{7.8cm}
\includegraphics{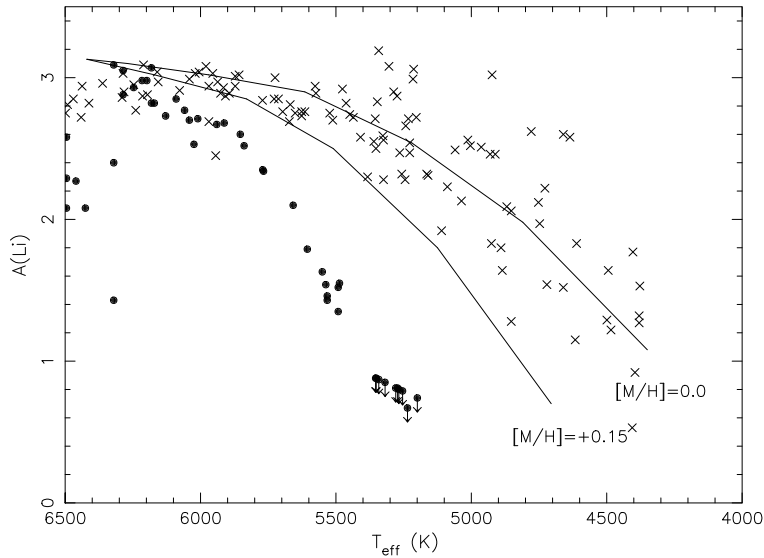}
\caption{Li abundances for members of the Pleiades (crosses) and Hyades
(dots). The solid lines are standard model predictions of PMS Li
depletion for two compositions (from Pinsonneault 1997).
}
\end{figure}

The two best studied open clusters are the Pleiades and Hyades, with
ages of $\simeq$100 and 600\,Myr, and consistently determined
spectroscopic iron abundances of [Fe/H]=$-0.034\pm0.024$ and
$+0.127\pm0.022$ (Boesgaard \& Friel 1990, Friel \& Boesgaard 1992).
Figure 1 shows Li abundances (determined using the same temperature
scale and curves of growth) for these clusters using data on {\em
single} stars gleaned from the sources in Table~1. Also shown are
standard Li depletion models for two mean metallicities. The general pattern
of Li depletion in the Pleiades is modelled reasonably well. The Hyades is
more metal rich than the Pleiades, so we expect more PMS Li depletion,
although not nearly as much as is observed. Two classes of solution can be put
forward to explain this discrepancy. (a) Swenson, Stringfellow \&
Faulkner (1990) show
that increasing interior opacities by modest amounts
could bring standard models into agreement with the Hyades data. Such
arguments do not explain why short-period, tidally locked {\em
binary systems} in the Hyades are much less Li depleted than
their single counterparts (Thorburn et al. 1993, Barrado y
Navascu\'{e}s \& Stauffer 1996).  (b) Extra mixing
whilst on the MS, driven by rotation and angular momentum loss seems
capable of providing the additional Li depletion with a natural
explanation for why the Li depletion in tidally locked binaries might
be different (Chaboyer et al. 1995).

Standard models also struggle to explain spreads in Li abundance among
late G and K-type Pleiades stars. The scatter appears to be correlated
with rotation, although a more detailed consideration ({\em e.g.}
Randich et al. 1998) shows that the correlation in both the Pleiades
and $\alpha$ Per clusters is driven largely by the fact that fast
rotating stars have suffered little Li depletion, whereas slowly
rotating stars can have either high or low Li abundances (see Figure
2). There are some indications that this dispersion may decrease again
at surface temperatures below 4500\,K (Jones et al. 1996).  One
interpretation would be to invoke non-standard mixing during the PMS
phase and the disk coupling paradigm for early angular momentum
evolution (Bouvier et al. 1997). Slow rotators might suffer little
extra mixing because they are born slow rotators and lose little
angular momentum, or they could be born as fast rotators and lose
considerable angular momentum by coupling to a long-lived circumstellar
disk and consequently undergo greater mixing and Li depletion. Stars
which are still fast rotators on the ZAMS would have been only briefly
coupled to a disk, would not have lost significant angular momentum and
suffered less internal mixing.  The problem with this explanation may
be that insufficient extra mixing associated with angular momentum can
take place on the PMS, and that fast rotators in the Pleiades have Li abundances
that lie above even the standard model predictions.

Adherents to the standard models could appeal to small metallicity
variations between cluster stars or to the possibility that atmospheric
inhomogeneities such as plages or starspots could cause a scatter in
the equivalent widths of \lii\ lines at a given $B-V$ value. This
latter explanation has been reviewed by Stuik, Bruls \& Rutten (1997),
who make a plausible case for considering such effects and point out
that the similarly formed \ki\,7699\AA\ line shows a nearly equivalent
scatter in Pleiades stars. As K abundance variations are not expected,
then the scatter in \ki\ equivalent widths at a given
colour means that it is premature to ascribe the {\em apparent} Li
abundance variation in late-type Pleiads to non-standard processes.

\begin{figure}
\vspace*{7.4cm}
\includegraphics{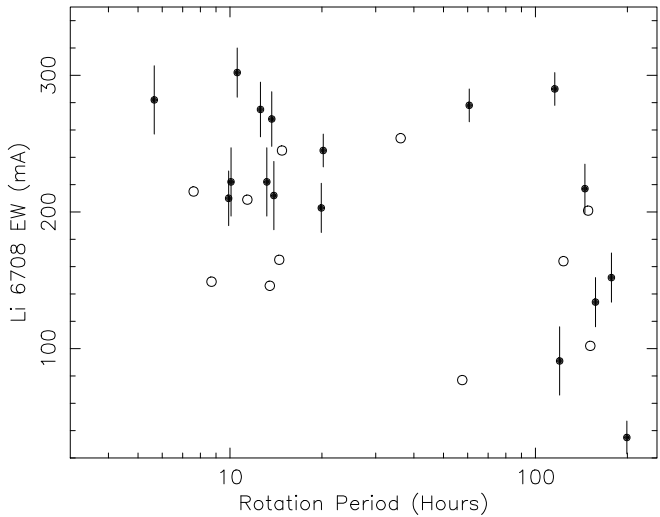}
\caption{The Li\,{\sc i}\,6708\AA\ line EWs for members of the Pleiades
(dots) and $\alpha$ Per (circles) as
a function of rotation period. Li measurements are from sources in
Table 1. A narrow colour range of $0.81<B-V<1.07$ was chosen for the
Pleiades and $0.88<B-V<1.13$ for $\alpha$ Per.
Rotation periods are from O'Dell et al. (1995, and references therein) 
and Krishnamurthi et al. (1998).}
\end{figure}

\section{Metallicity, age and Li depletion}

A natural question to ask is whether the Li depletion pattern in the
Hyades when it was younger, looked like that in the Pleiades now?
Standard models predict that the Hyades would look about the same as
they do now because all the depletion occurred during PMS evolution (for
$T_{\rm eff}\geq 5000$\,K).  Non-standard models predict a level of Li
depletion somewhere between the present day Hyades and Pleiades levels,
due to 500\,Myr of non-standard MS mixing.  Similarly, non-standard
models predict that if the Pleiades were aged to about 600\,Myr, the Li
depletion pattern should lie between the present day Pleiades and
Hyades because of reduced PMS Li depletion in the metal-poor Pleiades,
followed by somewhat less efficient MS Li depletion than in the Hyades
because of shallower CZs at a given $T_{\rm eff}$.  These are very
clear predictions. To test them simply requires Li abundance
measurements in the G and K stars of a cluster at the age of the
Hyades, but with the metallicity of the Pleiades, and vice-versa. These
data now exist in the form of Li abundances in the Blanco 1, and Coma
Berenices open clusters.

\subsection{Blanco 1}

Jeffries \& James (1999) present Li abundances for G and K stars in
Blanco 1, a young cluster (age 70\,Myr) with a spectroscopically
determined iron abundance of [Fe/H]=+0.14, when derived using the same
colour-$T_{\rm eff}$ scale as used for other young clusters.  Figure~3
presents the Li abundances of late-type stars in Blanco 1 compared with
the Pleiades and Hyades. Clearly the Blanco 1 Li abundances are
indistinguishable from those in the Pleiades and much higher than in
the Hyades.

These observations present problems for {\em both} standard and
non-standard Li depletion models. If the Hyades looked like Blanco 1 in
the past then non-standard MS mixing and Li depletion is clearly
indicated, because the stars in Blanco 1 should evolve to look like the
Hyades in $\sim500$\,Myr, offering useful empirical constraints on the
timescale for the mixing mechanisms. However, because non-standard
models predict extra depletion compared with the standard models, an
additional ingredient is required to explain why Blanco 1 has not
suffered significantly more initial PMS Li depletion than the Pleiades,
given it's higher metallicity.

\subsection{The Coma Berenices Open Cluster}

The sparse Coma Berenices open cluster (age 500\,Myr) has
[Fe/H]=$-0.052\pm0.026$, determined in a rigorously consistent way with
that of the Pleiades and Hyades values already quoted (Friel \&
Boesgaard 1992). Li abundances for G and K stars are presented by
Jeffries (1999) and supplemented with a few more observations by Ford
et al. (1999 - A\&A submitted).  The data for single stars are also
shown in Figure~3.  The Li depletion pattern for Coma Ber is very
similar to that in the Hyades, with perhaps a hint of less Li depletion
for stars cooler than 5700\,K.  Again, {\em both} standard and
non-standard models have problems explaining these observations. The
standard models would have that the Li depletion in Coma Ber, which
occurred during PMS evolution, should be similar to or even less than
that in the Pleiades. The extra depletion observed could be supplied by
non-standard mixing (on timescales that agree very well with the
Hyades-Blanco 1 comparison), but it is then hard to see why Coma Ber
and the Hyades should be so close at the present day, unless the PMS Li
depletion was not metallicity dependent and both clusters started out
on the ZAMS with similar depletion patterns -- as indicated by the
Pleiades and Blanco~1 datasets.

\begin{figure}
\vspace*{7.8cm}
\includegraphics{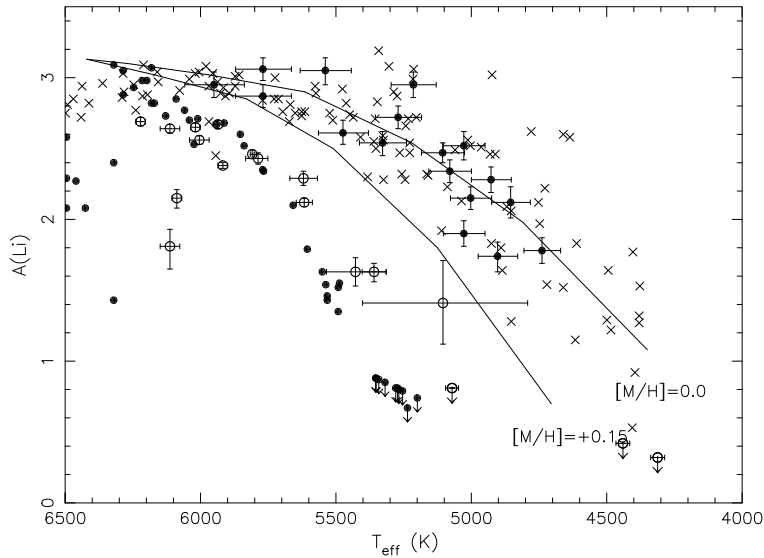}
\caption{Same as Figure 1 but this time data from the Blanco 1 (dots
with errors) and Coma Berenices (circles with errors) open clusters
have been added.
}
\end{figure}

\subsection{Other clusters}

To these two examples could be added Li abundance datasets for
IC 2391/2602, $\alpha$~Per, IC 4665 and NGC 2516 (see Table 1). These
clusters are either a
little younger or a little older than the Pleiades and probably have a wide
(albeit ill determined) range of metallicities. Yet the G and K stars
in these clusters have Li depletion patterns {\em very} close
to that in the Pleiades.
Similarly, Praesepe and NGC 6633 have ages close to that of the Hyades,
probably lower metallicities, yet show almost the same Li depletion
pattern as the Hyades. There is perhaps some evidence in NGC 6633 that
the K stars have not suffered quite as much depletion as in the Hyades,
but they are significantly more depleted than the Pleiades (Jeffries
1997). There are also clusters
with intermediate ages (NGC 1039, NGC 6475, 200-300\,Myr) which show
intermediate Li depletion patterns.

The global cluster dataset is clearly telling us that metallicity is
{\em not} an important parameter in determining the amount of PMS Li
depletion, which flatly contradicts the predictions of standard models.
Non-standard mixing processes acting during MS evolution are required
in order to rank the cluster Li depletion patterns according to age.
Their appear to be no significant exceptions to this trend.  The only
ways of rescuing the conventional view of standard models are to either
abandon the idea that one cluster is representative of clusters at the
same age and composition, or assume that [Fe/H] is not representative
of the overall metallicity of these clusters. Swenson et al. (1994)
have shown that abundances of elements such as O and Si are important
in determining CZ depth and PMS Li depletion. Detailed abundance
analyses of key clusters are required to check that we are not seeing
the effects of drastically non-solar abundance ratios, however this
explanation would seem to require an unlikely conspiracy of
circumstances, given the number of observed clusters.

For clusters with greater than solar metallicity, arriving on the ZAMS
with similar Li depletion patterns to the Pleiades, a mechanism is
indicated that severely reduces the predicted efficiency of Li
depletion on the PMS. This requirement can be extended to lower
metallicity clusters and is even more extreme if standard models
incorporating the full spectrum of turbulence convection model are
considered (Ventura et al. 1998). It has been suggested that structural
changes associated with rapid rotation might do this job (Mart\'{i}n \&
Claret 1996) and at the same time, explain the Li abundance scatter in
late-type Pleiades stars.  Recently, Mendes, D'Antona \& Mazzitelli
(1999) have shown that the effects of rapid rotation might actually be
in the opposite sense required and in any case, even the slow rotators
in Blanco 1 have similar Li abundances to analogous stars in the
Pleiades. Ventura et al. (1998) hypothesize that dynamo generated
magnetic fields could steepen the adiabatic temperature gradient
sufficiently to alter CZ properties and significantly diminish Li
depletion. Stronger magnetic fields and less Li depletion would be
expected in fast rotators, possibly matching observations in the
Pleiades, Blanco 1 and other ZAMS clusters.  At present this model is
very crude, but the work of Ventura et al. shows that the size of the
effect might certainly be enough to explain the lack of PMS Li
depletion and its near independence of metallicity.

\section{Older clusters}

As the case for non-standard Li depletion has been made convincingly
for younger clusters it is natural to ask how observations of older
clusters might delineate the mechanisms and timescales responsible for
the extra mixing. The Hyades-Blanco 1 and Pleiades-Coma Ber comparisons
indicate a Li depletion rate of about 300-500\,Myr per dex for ZAMS
K-stars, and perhaps a factor $\sim2-3$ slower in G-stars.  If the Sun were
taken as representative for a star of it's age, the depletion rate in
early G stars must average out to $\simeq2$\,dex of depletion in
4\,Gyr.

Li abundances in a good sample of old open clusters would constrain
these timescales. Unfortunately old open clusters are relatively rare
and tend to be distant. Furthermore, the K-stars have probably depleted
Li beyond detection (although strong upper limits would be
useful). Table~1 summarises the observational state of play. The best
studied old open cluster is the solar-age M67. The data presented in
Jones, Fischer \& Soderblom (1999) and Pasquini, Randich \& Pallavicini
(1997) show an order of magnitude scatter in the Li abundances of
solar-type stars at this age, and significant depletion with respect to
standard model PMS Li depletion predictions.  The solar Li abundance is
positioned towards the lower end of the distribution.

That Li is detected at all in 4.5\,Gyr old solar-type stars probably
indicates that Li depletion slows from an initially higher rate on the
ZAMS.  This would certainly be expected for mixing mechanisms that were
driven by a slowly declining rate of rotation and angular momentum
loss. Jones et al. (1999) ascribe the spread in Li abundances to
non-standard mixing in stars with a spread in initial ZAMS rotation
rates. The abundance spread must develop over several Gyr, because the
Pleiades and Hyades G stars show only marginal signs of this spread at younger
ages (Thorburn et al. 1993). 
The stars with initially higher rotation rates would then be
those with the lowest Li abundances in M67 and vice-versa (reversing
the trend seen in Pleiades K stars!). The circumstantial evidence for
this, is that the proportion of low and high Li abundances in M67
approximately matches the proportions of fast and slow rotators in the
Pleiades. 

This intriguing notion needs bolstering with measured rotation rates in
M67 (although rotation rates may well have converged to be
indistinguishable). If the scenario could be confirmed, then Jones et al. (1999)
speculate that the low Li abundance of the Sun indicates that is was
rapidly rotating on the ZAMS. This may still be premature because M67
has a slightly sub-solar metallicity. We lack the evidence to say by
how much metallicity affects non-standard mixing on long
timescales, but if higher metallicities enhance	MS Li depletion, then
the Sun may yet turn out to have a high Li abundance for its age.
This could be addressed by observations of several older
clusters and would be important in understanding how much prior
depletion has occurred in very metal-poor Population II stars.

\section{Conclusions}

I end by attempting to briefly answer the original questions in the abstract.
It is clear from the evidence reviewed that standard
stellar evolution models struggle to explain the patterns of Li
depletion seen in open clusters. Furthermore, observations of clusters
with different metallicities provide difficulties for current
non-standard models. There are strong indications that PMS Li depletion
is not as strong as predicted in either class of model. This has not
yet been widely recognized and hence explanations are so far rather speculative.

There are many pieces of evidence that non-standard mixing and Li
depletion are important during MS evolution. These include the
Hyades-Blanco 1 and Pleiades-Coma Ber comparisons, where the confusing
factor of metallicity dependent PMS Li depletion has been removed, the
general ordering of cluster Li depletion according to age and the
strong depletion seen among older clusters and the Sun. The timescales
for MS Li depletion are longer than PMS Li depletion timescales but are
still uncertain. The current observational evidence suggests that the
MS depletion timescales are shorter for K stars than G stars and may
get longer as stars spin down.

Metallicity appears not to play a great role in PMS Li depletion,
contradicting expectations. Abundance analyses
are required for O and Si to see whether CZ depth is affected
by non-solar abundance ratios, although the number of
clusters in the extant dataset makes this possibility unlikely.  If metallicity is
not important for PMS Li depletion, then one of the major uncertainties
in using Li abundances to date young stars is removed. The other is the
scatter in abundances seen at a given age, which inevitably introduces
uncertainties that can be well quantified by comparison with cluster
datasets. Thus although using Li abundances to age young stars might be
relatively inaccurate, depending on the spectral-type of star
considered, the uncertainties can at least be empirically determined.
It is very difficult however to date older stars using Li abundances
because (a) they also show a scatter in Li abundance that develops with
age, (b) stars cooler than G-type won't have detectable Li once older
than $\sim1$\,Gyr and (c) we still don't know whether metallicity
greatly affects the efficiency of MS Li depletion.

New observations could be made which would clarify a number of these
issues. Detailed abundance analyses could be performed for all the key
open clusters to check for non-solar abundance ratios.  Li abundance
measurements in several more old open clusters might betray any
metallicity dependence of MS depletion timescales.  Measuring rotation
periods in many more cluster stars with Li abundances, including the
slower rotators in older clusters where these measurements tend to be
much more difficult, would allow further investigation of how depletion
timescales depend on rotation rates.  The connection between Li
depletion and rotation in cool ZAMS stars is still far from resolved
and may yet turn out to be problems in our understanding of
inhomogeneous stellar atmospheres. In that respect, the connection
between rotation, surface inhomogeneities, \lii, and \ki\ equivalent
width spreads needs to be carefully investigated, possibly using
doppler tomographic techniques ({\em e.g.} Hussain, Unruh \&
Collier-Cameron 1998).

\acknowledgements

The author would like to thank the staff at the Isaac Newton Group of
Telescopes and the Anglo Australian Observatory for their assistance
during the course of several observing campaigns.

\end{document}